\documentclass[twocolumn,showpacs,preprintnumbers,amsmath,amssymb]{revtex4}

\usepackage{graphicx}
\usepackage{dcolumn}
\usepackage{bm}

\begin{document}

\preprint{APS/123-QED}

\title{Dynamics of Dirac-Born-Infeld dark energy interacting with dark matter}

\author{Chakkrit Kaeonikhom}
\affiliation{Department of Science, Chiang Mai Rajabhat University, Chiang Mai, 50300 Thailand}
 \email{chakkrit_kae@cmru.ac.th}
 
\author{Douglas Singleton}
\affiliation{Physics Department, CSU Fresno, Fresno, CA 93740 USA \\
and \\
Institut f{\"u}r Mathematik, Universit{\"a}t Potsdam
Am Neuen Palais 10, D-14469 Potsdam, Germany}%
 \email{dougs@csufresno.edu}

\author{Sergey V. Sushkov} 
\affiliation{Institute of Physics, Kazan Federal University, Kremlevskaya Street 18, Kazan 420008, Russia\\
and \\
Physics Department, CSU Fresno, Fresno, CA 93740 USA}
 \email{sergey_sushkov@mail.ru}

\author{N. Yongram}
 \altaffiliation{Corresponding author}
\affiliation{%
ThEP's CRL, NEP, The Institute of Fundamental Study (IF)\\
Naresuan\;University,\;Phitsanulok\;65000,\;Thailand\\
Thailand Center of Excellence in Physics\\
CHE, Ministry of Education, Bangkok 10400, Thailand\\
Physics Department, CSU Fresno, Fresno, CA 93740 USA}
 \email{nattapongy@nu.ac.th}

\date{\today}

\begin{abstract}
We study the dynamics of Dirac-Born-Infeld (DBI) dark energy interacting with dark matter. The DBI dark energy model
considered here has a scalar field with a nonstandard kinetic energy term, and has potential and brane tension
that are power-law functions. The new feature considered here is an interaction between the DBI dark energy and
dark matter through a phenomenological interaction between the DBI scalar field and the dark matter fluid.
We analyze two different types of interactions between the DBI scalar field
and the dark matter fluid. In particular we study the phase-space diagrams of and look for critical points of the phase space
that are both stable and lead to accelerated, late-time expansion. In general we find that the interaction
between the two dark components does not appear to give rise to late-time accelerated expansion. However, the interaction
can make the critical points in the phase space of the system stable. Whether such stabilization occurs or not
depends on the form of the interaction between the two dark components.   
\end{abstract}


\maketitle

\section{\label{Sec01}Introduction}

In recent years much of the effort in theoretical physics has gone into the study of the observed present accelerated 
expansion of the Universe first reported in \cite{Perlmutter_1999,Riess_1998} through 
observational data from Type Ia supernovae.
Subsequent work on Type Ia supernovae \cite{Hicken}, the cosmic microwave background \cite{Komatsu;2009}, 
and baryon acoustic oscillations \cite{Percival;2007}, all support the initial observations that the expansion of the
Universe is accelerating. This late-time acceleration of the Universe is driven by a fluid/field generically called
dark energy. Very little is known about dark energy. Within the context of string theory there is a model for the 
{\it early-time} accelerated expansion of the Universe associated with inflation. This string-theory-motivated model 
for inflation is called Dirac-Born-Infield (DBI) inflation
\cite{Silverstein_Tong;2004,Alishahiha_Silverstein_Tong;2004,Chen_prd;2005,Chen_jhep;2005,Shandera_Tye;2006, Gumjudpai;2004}, 
and it is driven by the open string sector through dynamical Dp-branes. 
DBI inflation is a special case of K-inflation models \cite{Armendariz;1999}. It was originally thought that 
DBI inflation models would yield large non-Gaussian perturbations which could be used to verify or 
falsify these models and by extension to test string theory \cite{Gumjudpai;2004,Chimento;2009}. 
However, subsequent work has shown that this may not be the case, and that the simplest DBI models
are effectively indistinguishable from standard field-theoretic slow-roll models of inflation \cite{Lidsey;2007}. 

In the present work we examine variants of these DBI models as a mechanism, not for the early-time 
acceleration of inflation, but for the observed late-time acceleration. Our DBI scalar field will play the role 
of dark energy. The action for our scalar DBI field is taken to have the form found in Ref. \cite{Chimento;2009}. 
\begin{equation}\label{eq01}
   S_{DBI} = -\int \mathrm{d}^{4}x \sqrt{-g}\left[ T(\phi)\sqrt{1-\frac{\dot{\phi}^{2}}{T}}+V(\phi)-T(\phi)\right] ~,
\end{equation}
where we have assumed that the scalar DBI field , $\phi$, is spatially homogeneous so that its spatial derivatives 
can be ignored. This is in accord with the fact that dark energy seems to be very homogeneously distributed.
Note that $\phi$ has a nonstandard kinetic energy term which yields a standard
kinetic energy term if one expands the square root to first order in $\frac{\dot{\phi}^{2}}{T}$. 
For a pure AdS$_{5}$ geometry with radius $R$, the warped tension $T(\phi)=\tau\phi^{4}$ is the D3-brane tension 
with $\tau=1/(g_{s}\tilde{\lambda})$, where $g_{s}$ is the string coupling, $\alpha^{\prime}$ is the inverse string tension, 
and $\tilde{\lambda}=R^{4}/\alpha^{\prime}$ is the 't Hooft coupling in the AdS/CFT correspondence. $V(\phi)$ is the 
potential arising from interactions with Ramond-Ramond fluxes or with other sectors \cite{Silverstein_Tong;2004}. 
Here we take the potential to be quadratic, $V(\phi)=m^{2}\phi^{2}$, and the associated D-brane is in the 
anti-de Sitter throat \cite{Guo_ Ohta;2008}. 

In a spatially flat Friedmann-Robertson-Walker (FRW) metric with scale factor $a(t)$, it can be shown that the 
energy density $\rho_{\phi}$ and the pressure $p_{\phi}$ of the DBI scalar field are given by
\begin{equation}
\rho_{\phi}=\frac{\gamma^{2}}{\gamma+1}\,\dot{\phi}^{2}+V(\phi) \quad\mathrm{and}\quad
p_{\phi}= \frac{\gamma}{\gamma+1}\,\dot{\phi}^{2}-V(\phi) , \label{02}
\end{equation}
where $\gamma$ has the form of a Lorentz boost factor,
\begin{equation}\label{eq03}
   \gamma\equiv\frac{1}{\sqrt{1-\frac{\dot{\phi}^{2}}{T(\phi)}}}.
\end{equation}
In this paper we analyze this system of a DBI dark energy field interacting with dark matter in terms of late-time scaling 
solutions. Such models are different from the original work in Ref. \cite{Copeland_Mizuno_Shaeri;2010} which studied the 
dynamics of a DBI field plus a perfect fluid  but with no interaction between the DBI field and the perfect fluid. 
In the course of our analysis of this model of DBI dark energy interacting with dark matter, we find that
for certain parameters there are late-time attractor solutions or fixed points in the phase space of the parameters. 

The organization of the rest of the paper is as follows. 
In Sec. II, we write down two possible phenomenological interactions between the DBI dark energy scalar field
with the dark matter fluid. In Sec. III, we analyze the autonomous equations in terms of the relevant variables of
DBI dark energy interacting with dark matter. We find the fixed points in the phase-space flow of
these variables. We also discuss the stability (i.e. stable fixed point, unstable fixed point or
saddle fixed point) of fixed points and determine if the fixed points correspond to late-time accelerated expansion
or not. Of particular importance will be stable fixed points which lead to late-time accelerated expansion.
Finally, we summarize our results in Sec. IV.

\section{DBI dark energy scalar field interacting with dark matter}

Cosmological evolution is thought to be largely dominated by dark energy and dark matter. Dark energy gives 
a gravitationally repulsive effect while dark matter is gravitationally attractive. Usually there is no interaction 
between these two components such as in the model in Ref.\cite{chaves-2007} where graded Lie algebras were
used to give a unified theory with both dark energy and dark matter, but without any interaction between 
these two components.

There has been work such as the two measure cosmological model of Ref. \cite{guendelman} where 
there is some interaction between the dark energy and dark matter components of the model. 
However, since the gravitational effects of dark energy and dark matter are opposite (i.e., gravitational
repulsion versus gravitational attraction) and since dark energy appears to be very homogeneously distributed,
while dark matter clumps around ordinary matter, one expects that any interaction between these two
dark components of the Universe would be weak. In this paper we will consider models where there is 
an interaction between dark energy and dark matter. The dark energy component will come 
from the DBI scalar field in Eq. \eqref{eq01} with energy density
$\rho _{\phi}$ and pressure $p_{\phi}$, and the dark matter component will come
from a fluid with an equation of state $w_{m}\equiv p_{m}/\rho_{m}=0$. 
Considering a spatially flat Friedman-Robertson-Walker background with 
scale factor $a(t)$, and allowing for creation/annihilation between the DBI scalar field and the dark matter
fluid at a rate $Q$, we can write down the equations for  $\rho_{\phi}$ and  $\rho_{m}$ as
\begin{eqnarray}
\dot{\rho}_{\phi}+3H\left(1+w_{\phi}\right)\rho_{\phi} &=& -Q\label{eq04}\\[0.5\baselineskip]
\dot{\rho}_{m}+3H\rho_{m} &=& +Q\label{eq05} ~.
\end{eqnarray}
Here $H\equiv \dot{a}/a$ is the Hubble rate with derivatives with respect to cosmological time, $t$, being indicated by 
a dot. The DBI scalar field and the dark matter create/decay into one another via the common creation/annihilation
rate $\pm Q$. $Q$ represents the interaction between these two fields. Although at this point this interaction is
generic, one can say that if $Q>0$ dark energy converts to dark matter, and if $Q<0$ dark matter
is converted to dark energy \cite{couplingmodels}.

Since there is no fundamental theory which specifies a coupling between dark energy and dark matter,
our coupling models will necessarily be phenomenological, although one might view some couplings as
more physical or more natural than others. In this paper we consider two types of coupling:
\begin{eqnarray}
\mathrm{Model ~ I}\quad  &Q&= \sqrt{\frac{2}{3}}\beta\rho_{m}\dot{\phi}\label{eq06},\\[0.5\baselineskip]
\mathrm{Model ~ II}\quad &Q&= \alpha H \rho_{m}\label{eq07}
\end{eqnarray}
where $\beta$ and $\alpha$ are dimensionless constants whose sign determines the direction of energy transfer. 
For positive values of the parameters $\alpha, \beta>0$ ($Q>0$) there is a transfer of energy from 
DBI dark energy to dark matter; for negative  values of the parameters $\alpha,\beta <0$ ($Q<0$) there is 
a transfer of energy from dark matter to DBI dark energy. 

The interaction given in Model I may be motivated within the context of scalar-tensor theories 
\cite{Wetterich;1995,Amendola;1999,Holden_Wards;2000} where similar interaction terms can be found. 
Generalizations of this model allow for $\beta = \beta(\phi)$ and more general forms of $V(\phi)$ (e.g., Refs. 
\cite{Huey_Wandelt;2006,Bean_Magueijo;2001}). Interactions of the form given by Model II have been considered 
in Ref. \cite{Zimdahl_ Pavon;2001} which used $Q/H =\alpha_{m}\rho_{m}+\alpha_{\phi}\rho_{\phi}$ and 
in Ref. \cite{Guo_ Ohta;2007}, which used $Q/H =\alpha\Omega_{\phi}$. 

The equation for the rate of change of the Hubble parameter is
\begin{equation}\label{eq08}
\dot{H}=-\frac{1}{2}\left[\left(1+w_{\phi}\right)\rho_{\phi}+\rho_{m}\right] 
\end{equation}
The Hubble parameter is subject to the constraint
\begin{equation}\label{eq09}
H^{2}=\frac{1}{3}(\rho_{\phi}+\rho_{m}) ~.
\end{equation}
In this work we use the units such that $8\pi G=1$, where $G$ is Newton's gravitational constant. 

We define the fractional density of the DBI dark energy and dark matter via $\Omega_{\phi}\equiv \rho_{\phi}/3H^{2}$ 
and $\Omega_{m}\equiv \rho_{m}/3H^{2} $, with the condition that $\Omega_{\phi}+\Omega_{m}=1$ which comes
from \eqref{eq09}. The modified Klein-Gordon equation, which follows from Eqs.\eqref{02}, \eqref{eq04}, and \eqref{eq08} 
and gives the evolution of the DBI scalar field,  takes the form
\begin{equation}\label{eq10}
\ddot{\phi}+\frac{3H}{\gamma^{2}}\dot{\phi}+\frac{V_{,\phi}}{\gamma^{3}}-
\frac{T_{,\phi}}{2T}\frac{(\gamma+2)(\gamma-1)}{(\gamma+1)\gamma}\,\dot{\phi}^{2}=-\frac{Q}{\gamma^{3}\dot{\phi}} ~,
\end{equation}
where $V_{,\phi}\equiv \mathrm{d}V(\phi)/\mathrm{d}\phi$ and $T_{,\phi}\equiv \mathrm{d}T(\phi)/\mathrm{d}\phi$. 
Eqs. \eqref{eq05}, \eqref{eq08}, and \eqref{eq10} give a closed system of equations that determines the dynamics of
the DBI dark energy scalar field, $\phi$, interacting with dark matter.

In order to find the fixed points of this system and to study the late-time attractor 
behavior of these two models we introduce the following set of dimensionless variables 
similar to those used in Ref. \cite{Copeland_Liddle;1998}:
\begin{equation}\label{eq11}
x\equiv\frac{\gamma\dot{\phi}}{\sqrt{3(\gamma+1)}H},\quad y\equiv\frac{\sqrt{V}}{\sqrt{3}H}.
\end{equation}

With $x$ and $y$ defined in this way we can recover the variables for the scalar field model 
originally proposed in Ref. \cite{Copeland_Liddle;1998} by taking the limit $\gamma\to 1$.
The variable $x$ roughly corresponds to the the kinetic energy of the DBI field,
while $y$ roughly corresponds to the potential energy of the DBI field. In addition to these dynamical 
variables $x$ and $y$ we introduce a third variable $\tilde{\gamma}=1/\gamma$ which is connected 
with the brane tension $T(\phi)$. Taking the inverse of $\gamma$ makes the
final equations more compact. Thus we have exchanged the three variables $\phi$, $V (\phi)$, and
$T(\phi)$ for $x$, $y$, and $\tilde{\gamma}$.  

In terms of these variables, the Friedmann constraint from Eq. \eqref{eq08} can be expressed as
\begin{equation}\label{eq12}
x^{2}+y^{2}+\Omega_{m}= \Omega_{\phi}+\Omega_{m}= 1.
\end{equation}
The equation of state of the DBI dark energy is given by
\begin{equation}\label{eq13}
w_{\phi} = \frac{\rho _{\phi}}{p_{\phi}} = \frac{\tilde{\gamma} x^{2}-y^{2}}{x^{2}+y^{2}}.
\end{equation}
Finally, we introduce two sets of variables related to the potential, $V(\phi)$, and the 
brane tension, $T(\phi)$. The first set, $\lambda_{1}$ and $\lambda_{2}$, are defined as
\begin{equation}
\lambda_{1}\equiv-\frac{V_{,\phi}}{V},\qquad \lambda_{2}\equiv-\frac{T_{,\phi}}{T} ~.\label{eq14}
\end{equation}
The second set of variables, $\tilde{\lambda}_{1}$ and $\tilde{\lambda}_{2}$, are given by
\begin{equation}
\tilde{\lambda}_{1}\equiv-\frac{V_{,\phi}}{T^{-1/2}V^{3/2}},\qquad 
\tilde{\lambda}_{2}\equiv-\frac{T_{,\phi}}{T^{1/2}V^{1/2}} \label{eq15}
\end{equation}
The relationship between the two sets of variables \eqref{eq14} and \eqref{eq15} is given by
\begin{equation}
\lambda_{1}=\sqrt{\frac{(1-\tilde{\gamma})}{\tilde{\gamma}}}\frac{y}{x}\tilde{\lambda}_{1},
\qquad \lambda_{2}=\sqrt{\frac{(1-\tilde{\gamma})}{\tilde{\gamma}}}\frac{y}{x}\tilde{\lambda}_{2}.\label{eq16}
\end{equation}
Combining the above definitions, the evolution equations for $x$, $y$ and ${\tilde \gamma}$ can be written 
as the following autonomous system:
\begin{eqnarray}
\frac{\mathrm{d}x}{\mathrm{d}N}=&&\frac{ \tilde{\lambda}_{1} y^{3}\sqrt{3(1-\tilde{\gamma}^{2})}}{2x}
-\frac{Q}{6 x H^{3}}+\frac{3x}{2}\left\{\tilde{\gamma}(x^{2}-1)-y^{2}\right\}\nonumber\\[0.5\baselineskip]
&&{}\label{eq17}\\[0.5\baselineskip]
\frac{\mathrm{d}y}{\mathrm{d}N}=&& -\frac{ \tilde{\lambda}_{1} y^{2}\sqrt{3(1-\tilde{\gamma}^{2})}}{2}
+\frac{3y}{2}\left\{1+\tilde{\gamma}x^{2}-y^{2}\right\}\label{eq18}\\[0.5\baselineskip]
\frac{\mathrm{d}\tilde{\gamma}}{\mathrm{d}N}=&&\frac{\tilde{\gamma}(1-\tilde{\gamma}^{2})}{\sqrt{(1+\tilde{\gamma})}}
\!\!\left[3\sqrt{(1+\tilde{\gamma})}\!\!-\!\!\sqrt{3(1-\tilde{\gamma})}\frac{y}{x^{2}}\!\left[ \tilde{\lambda}_{1} y^{2}
+\tilde{\lambda}_{2} x^{2}\right]\right.\nonumber\\[0.5\baselineskip]
&&\qquad\qquad\quad\quad\left.+\frac{Q}{x^{2}\sqrt{3(1+\tilde{\gamma})}H^{3}}\!\right]\label{eq19}
\end{eqnarray}
where $N\equiv\ln{a}$ and $\frac{\mathrm{d}\,\,}{\mathrm{d}N}=\frac{1}{H}\frac{\mathrm{d}\,\,}{\mathrm{d}t}$. 
There are also two equations for $\lambda_1$ and $\lambda_2$ 
\begin{eqnarray}
\frac{\mathrm{d} \lambda_{1}}{\mathrm{d}N}=&&-x  \lambda_{1}^{2}\sqrt{3\tilde{\gamma}(1+\tilde{\gamma})}
\left\{ \frac{VV_{,\,\phi\phi}}{V^{2}_{,\,\phi}}-1\right\},\label{eq20}\\[0.5\baselineskip]
\frac{\mathrm{d} \lambda_{2}}{\mathrm{d}N}=&&-x  \lambda_{2}^{2}\sqrt{3\tilde{\gamma}(1+\tilde{\gamma})}
\left\{\frac{TT_{,\,\phi\phi}}{T^{2}_{,\,\phi}}-1\right\} \; .\label{eq21}
\end{eqnarray}
Since $\mathrm{d}\lambda_{1}/\mathrm{d}N$ and $\mathrm{d}\lambda_{2}/\mathrm{d}N$ can be expressed in
terms of $\mathrm{d}x/\mathrm{d}N$, $\mathrm{d}y/\mathrm{d}N$, $\mathrm{d}\tilde{\gamma}/\mathrm{d}N$ 
from Eqs. \eqref{eq17} - \eqref{eq19}, thus equations Eqs. \eqref{eq20} and \eqref{eq21} are not
independent equations and we only need to solve three equations for $x$, $y$, and ${\tilde \gamma}$. 
The time evolution equation for $H$ given in Eq. \eqref{eq08} can be rewritten by
differentiating the Hubble parameter with respect to $N$, which yields   
\begin{equation}
\frac{1}{H}\frac{\mathrm{d}H}{\mathrm{d}N}=-\frac{3}{2}\left\{1+\tilde{\gamma}x^{2}-y^{2}\right\}.\label{eq22}
\end{equation}
The total effective equation of state for the DBI scalar field plus dark matter can be written as 
\begin{equation}\label{eq23}
w_{\mathrm{eff}}= \frac{p_{\phi}+p_{m}}{\rho_{\phi}+\rho_{m}} = \tilde{\gamma}x^{2}-y^{2}.
\end{equation}

In the next section we will investigate the two different models, given by Eqs.
\eqref{eq06} and \eqref{eq07}, near critical points ($x_{c}$, $y_{c}$). 
Near a critical point the scale factor of the FRW space-time takes the form
\begin{equation}\label{eq24a}
a\propto t^{2/3(1+\tilde{\gamma}x^{2}_{c}-y^{2}_{c})} \; .
\end{equation}
In order to have accelerated expansion (i.e. $\ddot{a}>0$) the above equation
requires that $w_{\mathrm{eff}} = \tilde{\gamma}x^{2}_{c}-y^{2}_{c}<-\frac{1}{3}$.
Combining Eq. \eqref{eq22} with Eq. \eqref{eq23}, and recalling that
$\frac{\mathrm{d}\,\,}{\mathrm{d}N}=\frac{1}{H}\frac{\mathrm{d}\,\,}{\mathrm{d}t}$,
the Hubble parameter evolution equation becomes
\begin{equation}\label{eq24}
\frac{\dot{H}}{H^{2}}=-\frac{3(1+w_{\mathrm{eff}})}{2}.
\end{equation}
The energy balance equations \eqref{eq04} and \eqref{eq05} for Model I and Model II 
are independent of $H$ when expressed in terms of the variables $x(N)$ and $y(N)$, where $N = \ln a$. 
Thus the Hubble parameter evolution equation \eqref{eq24} is not needed for these particular 
interacting models, and the phase space of both models is a two-dimensional phase space involving
$x$ and  $y$.

\section{Critical points and stability analysis}

In this section we find the critical or fixed points of the autonomous system Eqs.
\eqref{eq17} - \eqref{eq19} and perform a stability analysis of these fixed points. We are looking for 
the late-time attractor structure of this system of a DBI scalar field interacting with dark matter via
an energy exchange given by $Q$. The fixed points for Eqs. \eqref{eq17} - \eqref{eq19} are found by setting
$dx/dN = dy/dN = d{\tilde \gamma}/dN =0$ and solving the resulting three algebraic equations 
for the critical $x_c, y_c,$ and ${\tilde \gamma}_c$. Additionally, we will focus on the case when
${\tilde \gamma}_c =0$ [from Eq. \eqref{eq03} this implies $\gamma = \infty$] or 
${\tilde \gamma}_c =1$ [from Eq. \eqref{eq03} this implies $\gamma = 1$].  Thus 
${\tilde \gamma}$ is constant and the autonomous system reduces to only two dynamical variables: $x$ and $y$.

After finding the fixed points we study their stability with respect to small perturbations,  
$\delta x$ and $\delta y$, about the critical points $x_{c}$, $y_{c}$. Explicitly, these
take the form 
\begin{equation}
x=x_{c}+\delta x,\quad y=y_{c}+\delta y \;. \label{eq25}
\end{equation} 
Substituting Eq. \eqref{eq25} into Eqs. \eqref{eq17} and \eqref{eq18}, and keeping terms up to 
first order in $\delta x$ and $\delta y$, leads to a system of first-order 
differential equations of the form 
\begin{equation}\label{eq26}
\frac{\mathrm{d}}{\mathrm{d} N}\left( {\begin{array}{*{20}{c}}
   {\delta x}  \\
   {\delta y}  \\
\end{array}} \right)=\mathcal{M}\left( {\begin{array}{*{20}{c}}
   {\delta x}  \\
   {\delta y}  \\
\end{array}} \right) \:,   
\end{equation}
where $\mathcal{M}$ is a $2 \times 2$ matrix that depends on $x_{c}$ and $y_{c}$. 
To study the stability around the fixed points one calculates the eigenvalues of
$\mathcal{M}$. We denote these eigenvalues as $\mu_{1}$ and $\mu_{2}$, and for every critical 
point $(x_c, y_c)$ there is an associated eigenvalue pair $(\mu_1 , \mu_2)$. 
The stability of the critical point $(x_c, y_c)$ is then determined by its associated 
eigenvalue pair $(\mu_1 , \mu_2)$ in the following way: (i) if $\mu_{1}< 0$ and  $\mu_{2}< 0$, then the
critical point is stable; (ii) if $\mu_{1}> 0$ and $\mu_{2}> 0$, then the critical point is unstable; 
(iii) if $\mu_{1}<0$ and $\mu_{2}> 0$ or $\mu_{1}> 0$ and $\mu_{2}< 0$, then one has a saddle point;
(iv) if the determinant of the matrix $\mathcal{M}$ is negative and the real parts of $\mu_{1}$ and $\mu_{2}$ 
are negative, then one has a limit cycle. For our two models, Model I and Model II, all the fixed points
fall into cases (i), (ii), or (iii). None of the fixed point we found are limit cycles. 

\subsection{Interacting Model~I $Q=\sqrt{\frac{2}{3}}\beta\rho_{m}\dot{\phi}$} \label{subsec3.1.1}

Recalling that we are taking ${\tilde \gamma}$ to be a nondynamical constant with a value of $0$ or
$1$ our autonomous system \eqref{eq17} - \eqref{eq18} for Model I \eqref{eq06} becomes 
 \begin{eqnarray}
&&\frac{\mathrm{d}x}{\mathrm{d}N}=\frac{ \tilde{\lambda}_{1} y^{3}\sqrt{3(1-\tilde{\gamma}^{2})}}{2x}
-\frac{\beta\sqrt{2\tilde{\gamma}(1+\tilde{\gamma})}[1-x^{2}-y^{2}]}{2}\nonumber\\[0.5\baselineskip]
&&\qquad+\frac{3x}{2}\left\{\tilde{\gamma}\left(x^{2}-1\right)-y^{2}\right\}\label{eq27}\\[0.5\baselineskip]
&&\frac{\mathrm{d}y}{\mathrm{d}N}= -\frac{ \tilde{\lambda}_{1} y^{2}\sqrt{3(1-\tilde{\gamma}^{2})}}{2}
+\frac{3y}{2}\left\{1+\tilde{\gamma}x^{2}-y^{2}\right\}\label{eq28}
\end{eqnarray}
The dynamics of this autonomous system is determined by the parameters $\beta$ and $\tilde{\lambda}_{1}$.

The fixed points for Eqs. \eqref{eq27} and \eqref{eq28} are obtained by setting $dx/dN = 0$ and $dy/dN = 0$ and
solving the resulting algebraic equations for $x_c$ and $y_c$. There are six fixed points and these
are presented in the first two columns of Table \ref{tab01}.

\begin{table*}
\caption{\label{tab01} The fixed points for Model I with $Q=\sqrt{\frac{2}{3}}\beta\rho_{m}\dot{\phi}$.} 
\begin{ruledtabular}
\begin{tabular}{cccccccc}                             
Fixed point &$x_c$&$y_c$&$\tilde{\gamma}$&$\Omega_{\phi}$&$w_{\phi}$&$w_{\mathrm{eff}}$\\ 
\hline
(a1) &$-\sqrt{\frac{\tilde{\lambda}_{1}\left(\sqrt{\tilde{\lambda}^{2}_{1}+12}-\tilde{\lambda}_{1}\right)}{6}}$ &$\frac{\sqrt{\tilde{\lambda}^{2}_{1}+12}-\tilde{\lambda}_{1}}{2\sqrt{3}}$& $0$ &$1$   &$-\frac{[\sqrt{\tilde{\lambda}^{2}_{1}+12}-\tilde{\lambda}_{1}]^{2}}{12}$ & $-\frac{[\sqrt{\tilde{\lambda}^{2}_{1}+12}-\tilde{\lambda}_{1}]^{2}}{12}$ \\
(a2) &$\sqrt{\frac{\tilde{\lambda}_{1}\left(\sqrt{\tilde{\lambda}^{2}_{1}+12}-\tilde{\lambda}_{1}\right)}{6}}$ &$\frac{\sqrt{\tilde{\lambda}^{2}_{1}+12}-\tilde{\lambda}_{1}}{2\sqrt{3}}$& $0$ &$1$   &$-\frac{[\sqrt{\tilde{\lambda}^{2}_{1}+12}-\tilde{\lambda}_{1}]^{2}}{12}$ & $-\frac{[\sqrt{\tilde{\lambda}^{2}_{1}+12}-\tilde{\lambda}_{1}]^{2}}{12}$ \\
(b1) &$-1$ &$0$&$1$&$1$  &$1$ & $1$ \\
(b2) &$0$ &$1$&$1$&$1$  &$-1$ & $-1$ \\
(b3) &$1$ &0&$1$&$1$  &$1$ & $1$ \\
(b4) &$-\frac{2\beta}{3}$ &0&$1$&$\frac{4\beta^{2}}{9}$ & $1$ &$\frac{4\beta^{2}}{9}$ \\
\end{tabular}
\end{ruledtabular}
\end{table*}

We call the fixed points (a1)-(a2) ``ultrarelativistic'' fixed points
since for them the Lorentz factor $\gamma$ [Eq. \eqref{eq03}] tends to infinity 
(which implies that $\tilde{\gamma}=0$). There are also an infinite number of ``trivial" fixed points
for which ${\tilde \gamma}_c = y_c = 0$ and for which $x_c$ is arbitrary within the range 
constrained by $0 < x_c < \sqrt{1-\Omega_{m}}$ ($\Omega_{m}$ is the previously defined fractional density
of dark matter). These unstable, ``trivial" critical points are shown along the $y=0$ axis 
in Fig. \ref{fig01}. The four other critical points (b1)-(b4) listed in Table \ref{tab01} 
are ``standard" fixed points since for these fixed points the Lorentz factor of
Eq. \eqref{eq03} equals $1$ (so that $\tilde{\gamma}=1$), and the DBI field will mimic the behavior of 
a canonical scalar field.    

\subsubsection{{\bf Stability of the fixed points in Model I}}\label{subsec3.1.2}  

For each of the six fixed points listed in Table \ref{tab01} we found the
eigenvalues $\mu_{1}$ and $\mu_{2}$ of the matrix $\mathcal{M}$ in Eq. \eqref{eq26}. The results
for each point are listed below along with whether the point is stable, unstable, or a saddle point 
\begin{itemize}
  \item Point (a1):\\
          $\mu_{1}=-\frac{1}{4}\left(\tilde{\lambda}_{1}-\sqrt{\tilde{\lambda}^{2}_{1}+12}\right)^{2}$ ,\\  
    $\mu_{2}=\frac{1}{4}\left(-\tilde{\lambda}^{2}_{1}+\tilde{\lambda}_{1}\sqrt{\tilde{\lambda}^{2}_{1}+12}-12\right)$. \\
This point is stable for all values of $\tilde{\lambda}_{1}$.    
  \item Point (a2):\\
          $\mu_{1}=-\frac{1}{4}\left(\tilde{\lambda}_{1}-\sqrt{\tilde{\lambda}^{2}_{1}+12}\right)^{2}$,\\ 
$\mu_{2}=\frac{1}{4}\left(-\tilde{\lambda}^{2}_{1}+\tilde{\lambda}_{1}\sqrt{\tilde{\lambda}^{2}_{1}+12}-12\right)$. \\
This point is stable for all values of $\tilde{\lambda}_{1}$.
  \item Point (b1):\\
          $\mu_{1}=3$, $\mu_{2}=3-2\beta$. \\
This point is a saddle point for $\beta > \frac{3}{2}$ and is unstable for $\beta < \frac{3}{2}$.        
  \item Point (b2):\\
          $\mu_{1}=-3$, $\mu_{2}=-3$. \\
This point is stable for all values of $\beta$ and $\tilde{\lambda}_{1}$.
  \item Point (b3):\\
          $\mu_{1}=3$, $\mu_{2}=2\beta+3$. \\
This point is unstable for $\beta > -\frac{3}{2}$ and is a saddle point for $\beta > -\frac{3}{2}$.
 \item Point (b4):\\
         $\mu_{1}=\frac{1}{6}\left(4\beta^{2}-9\right)$, $\mu_{2}=\frac{1}{6}\left(4\beta^{2}+9\right)$. \\
This point is unstable for $\beta > \frac{3}{2}$ or $\beta < -\frac{3}{2}$ and is a saddle point for 
-$\frac{3}{2} < \beta < \frac{3}{2}$.
\end{itemize}

\begin{table*}
\caption{\label{tab02} The conditions for stability, acceleration and existence of the fixed point for Model I
in terms of the parameters $\beta$ and $\tilde{\lambda}_{1}$.} 
\begin{ruledtabular}
\begin{tabular}{llll}                 
Fixed point &Stability &Acceleration&Existence\\
\hline
(a1) &Stable node for all values of  $\tilde{\lambda}_{1}$   &$\tilde{\lambda}_{1}<2$  &all $\beta$, $\tilde{\lambda}_{1}$\\
(a2)  &Stable node for all values of  $\tilde{\lambda}_{1}$   &$\tilde{\lambda}_{1}<2$ &all $\beta$, $\tilde{\lambda}_{1}$\\
(b1) &Saddle point for $\beta>\frac{3}{2}$ & No & $\beta>0$\\
{}    &Unstable node for $\beta<\frac{3}{2}$ &{}&{}\\
(b2) &Stable node for all values of $\beta$, $\tilde{\lambda}_{1}$ &Yes&all $\beta$, $\tilde{\lambda}_{1}$\\
(b3) &Saddle point for $\beta<-\frac{3}{2}$ &No&all $\beta$, $\tilde{\lambda}_{1}$\\
{}    &Unstable node for $\beta>-\frac{3}{2}$ &{}&{}\\
(b4) &Saddle point for $-\frac{3}{2}<\beta<\frac{3}{2}$ &No&$\beta<0$\\
{}    &Unstable node for $\beta<-\frac{3}{2}$ or $\beta>\frac{3}{2}$&{}&{}\\
\end{tabular}
\end{ruledtabular}
\end{table*} 

\subsubsection{{\bf Late-time behavior for model I}}\label{subsec3.1.3}
 
In this subsection we investigate the late-time behavior of the scale factor $a(t)$. We are interested 
in whether $a(t)$ is accelerating, which means that the total effective equation of state parameter for this
model should satisfy $w_{\mathrm{eff}} < -\frac{1}{3}$. The six critical points from Table \ref{tab01} are listed in
Table \ref{tab02} with the conditions under which they are stable, unstable, or saddle points (the second
column), the conditions (if any) under which the critical point leads to accelerated expansion (the third 
column), and the conditions on $\beta, {\tilde \lambda}_1$ for the critical point to exist (the fourth column). 

\begin{figure}
\centering
\includegraphics[width=0.5\textwidth]{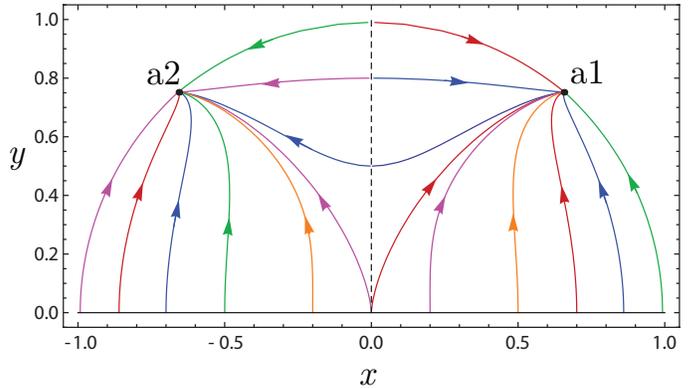}
\caption{\label{fig01}Model~I: The $x-y$ phase plane for the DBI dark energy interacting with dark matter. 
We have taken the parameters as $\beta=1$, $\tilde{\lambda}_{1}=1$, and $\tilde{\gamma}=0$. There are two
stable, critical points: a(1) and a(2). Both points lead to accelerated expansion. The dotted line is $x=0$, 
which is the singularity value of the nonallowed region.}
\end{figure}

Table \ref{tab02} shows that for Model I the two critical points (a1) and (a2) are stable for
all values of $\tilde{\lambda}_{1}$ since $\mu_{1}, \mu_{2} <0$. Further, these two points lead to 
accelerated expansion (i.e., $w_{\mathrm{eff}} <-\frac{1}{3}$) if $\tilde{\lambda}_{1} < 2$.  
For these reasons these two attractors are of interest in explaining the observed late-time accelerated 
expansion of the Universe. The phase-space flow for these two points is shown in Fig. \ref{fig01}. 

\begin{figure}
\centering
\includegraphics[width=0.5\textwidth]{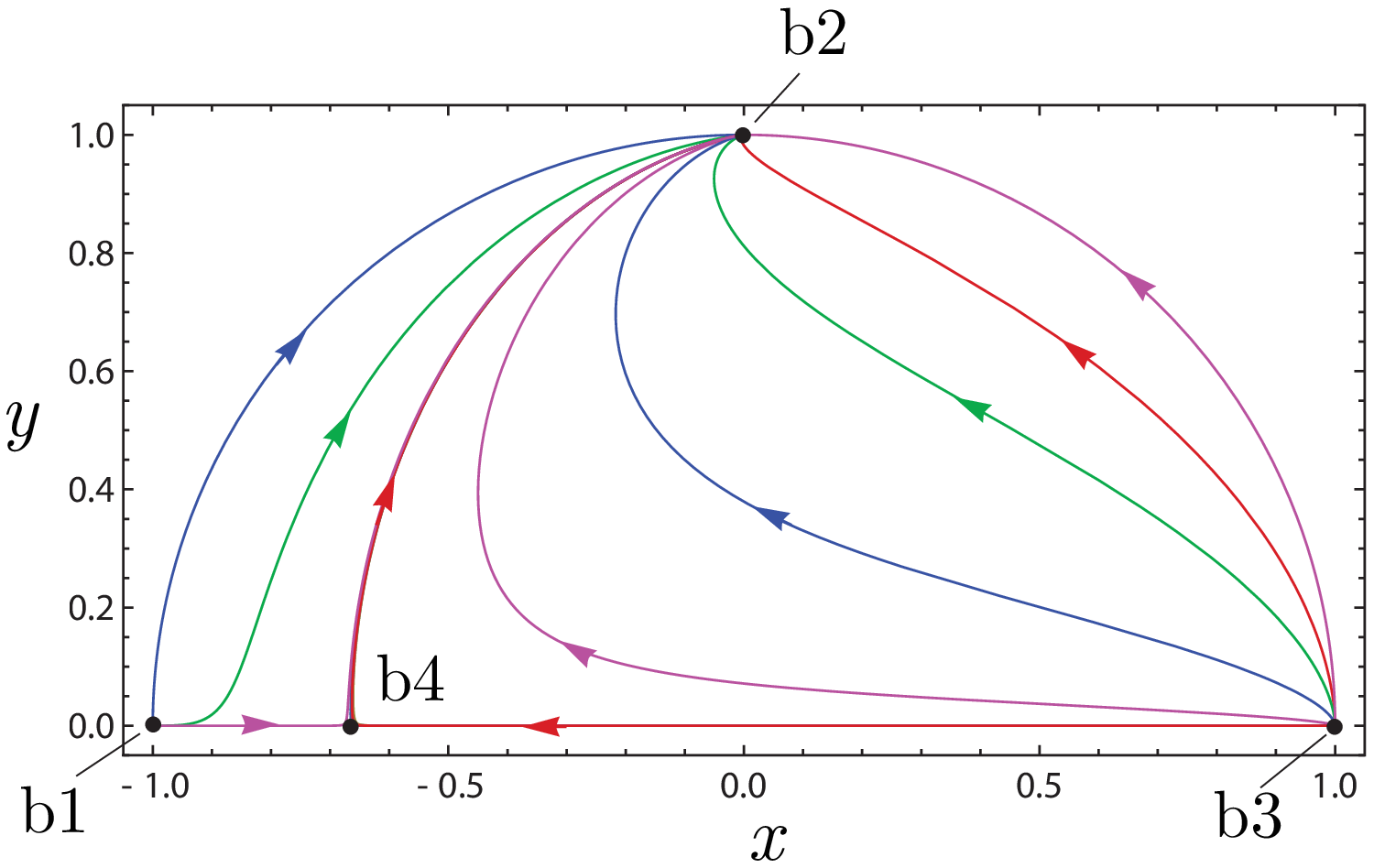}
\caption{\label{fig02}Model I: The $x-y$ phase plane for DBI dark energy interacting with dark matter 
with the parameters $\beta=1$, $\tilde{\lambda}_{1}=1$ and $\tilde{\gamma}=1$. The point $(b2)$ is a stable
critical point which leads to late-time accelerated expansion. }
\end{figure}

\begin{figure}
\centering
\includegraphics[width=0.5\textwidth]{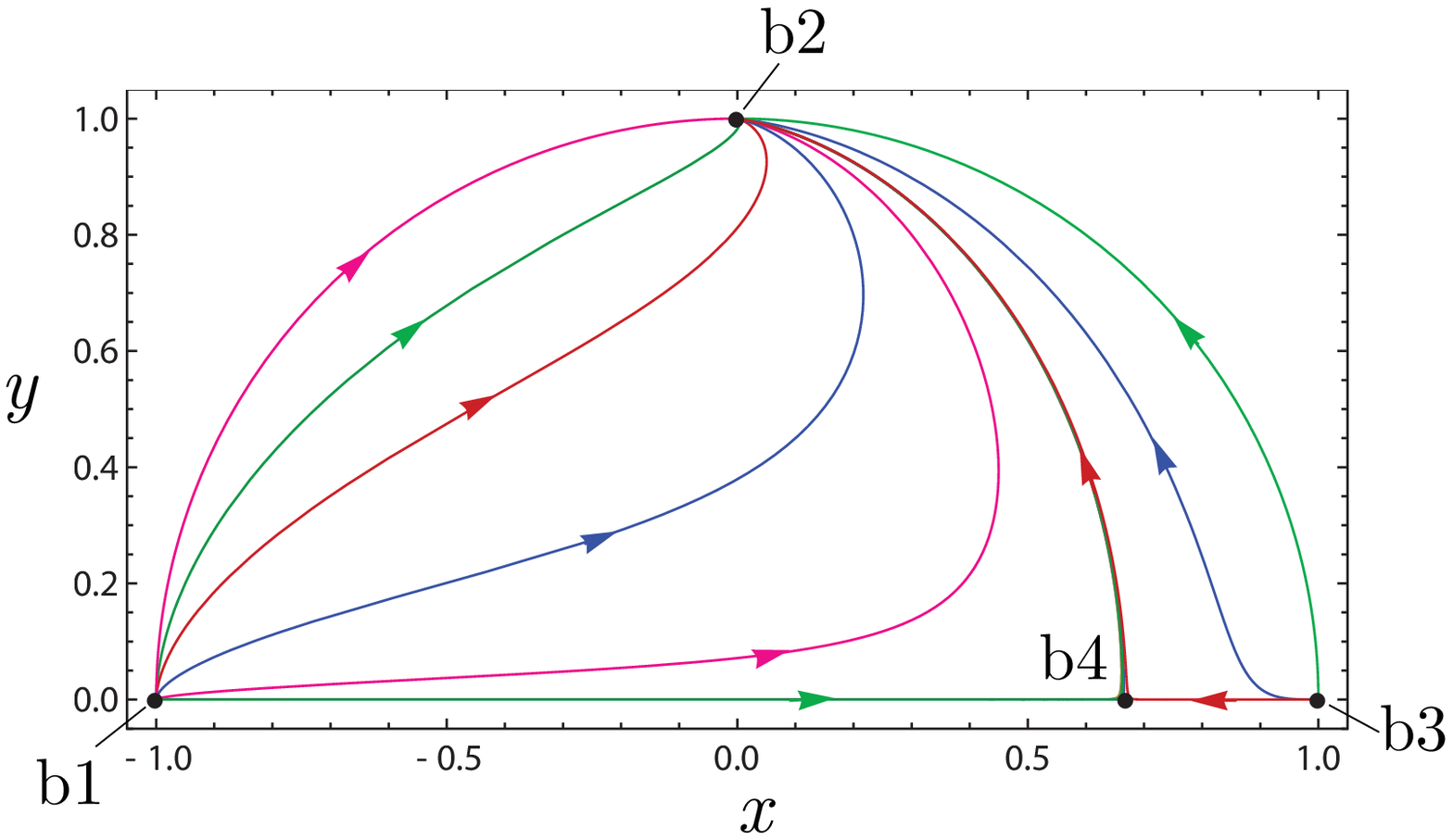}
\caption{\label{fig03}Model I: The $x-y$ phase plane for DBI dark energy interacting with dark matter 
with the parameters $\beta=-1$, $\tilde{\lambda}_{1}=1$ and $\tilde{\gamma}=1$. The point $(b2)$ is a stable
critical point which leads to late-time accelerated expansion.}
\end{figure}

The point (b2) is a stable critical point (i.e., $\mu_{1}, \mu_{2} <0$) for all 
values of $\beta$ and $\tilde{\lambda}_{1}$.  Additionally the point (b2) gives accelerated expansion
for the Universe (i.e., $w_{\mathrm{eff}} <-\frac{1}{3}$) for all values of $\tilde{\lambda}_{1}$ and $\beta$.
The $x-y$ phase-space behavior of the critical point (b2) is shown in Figs. \ref{fig02} and \ref{fig03}.
The remaining critical points -(b1), (b3), and (b4) - are of less interest phenomenologically since they do
not lead to accelerated expansion. Thus, there are three fixed points - (a1), (a2), and (b2) - 
which are stable and lead to late-time accelerated expansion. However, in regard to the existence of these
fixed points or their stability the specific value of $\beta$ (the parameter which characterizes the
coupling between dark energy and dark matter) plays no role. All these fixed points would exist and
be stable even if $\beta =0$, i.e., even in the absence of coupling between dark energy and dark matter.
Thus for Model I the overall conclusion is that this coupling does not play a significant role in 
the late-time behavior of the system.

\subsection{Interacting Model II: $Q=\alpha H\rho_{m}$}

The autonomous system for the variables $x$ and $y$ is now
 \begin{eqnarray}
&&\frac{\mathrm{d}x}{\mathrm{d}N}= \frac{ \tilde{\lambda}_{1} y^{3}\sqrt{3(1-\tilde{\gamma}^{2})}}{2x}-
\frac{\alpha}{2x}[1-x^{2}-y^{2}]\nonumber\\[0.5\baselineskip]
&&\qquad+\frac{3x}{2}\left\{\tilde{\gamma}\left(x^{2}-1\right)-y^{2}\right\}\label{eq30}\\[0.5\baselineskip]
&&\frac{\mathrm{d}y}{\mathrm{d}N}= -\frac{ \tilde{\lambda}_{1} y^{2}\sqrt{3(1-\tilde{\gamma}^{2})}}{2}
+\frac{3y}{2}\left\{1+\tilde{\gamma}x^{2}-y^{2}\right\}\label{eq31}
\end{eqnarray}

The fixed points are again obtained by setting $dx/dN = 0$ and $dy/dN = 0$ in Eqs. \eqref{eq30}
and \eqref{eq31}. For Model II there are eight fixed points - (d1) to (d4) and (e1) to (e4) - 
and these are listed in Table \ref{tab03}.

\begin{table*}
\caption{\label{tab03} The fixed points for Model II} 

\begin{ruledtabular}
\begin{tabular}{ccccccc}                            
Fixed point &$x_c$&$y_c$&$\tilde{\gamma}$&$\Omega_{\phi}$&$w_{\phi}$&$w_{\mathrm{eff}}$\\ 
\hline
(d1) &$-1$ &0&$0$&$1$  &$0$ & $0$ \\
(d2) &$1$ &0&$0$&$1$  &$0$ & $0$ \\
(d3) &$-\sqrt{\frac{\tilde{\lambda}_{1}\left(\sqrt{\tilde{\lambda}^{2}_{1}+12}-\tilde{\lambda}_{1}\right)}{6}}$ &$\frac{\sqrt{\tilde{\lambda}^{2}_{1}+12}-\tilde{\lambda}_{1}}{2\sqrt{3}}$& $0$ &$1$   &$-\frac{[\sqrt{\tilde{\lambda}^{2}_{1}+12}-\tilde{\lambda}_{1}]^{2}}{12}$ & $-\frac{[\sqrt{\tilde{\lambda}^{2}_{1}+12}-\tilde{\lambda}_{1}]^{2}}{12}$ \\
(d4) &$\sqrt{\frac{\tilde{\lambda}_{1}\left(\sqrt{\tilde{\lambda}^{2}_{1}+12}-\tilde{\lambda}_{1}\right)}{6}}$ &$\frac{\sqrt{\tilde{\lambda}^{2}_{1}+12}-\tilde{\lambda}_{1}}{2\sqrt{3}}$& $0$ &$1$   &$-\frac{[\sqrt{\tilde{\lambda}^{2}_{1}+12}-\tilde{\lambda}_{1}]^{2}}{12}$ & $-\frac{[\sqrt{\tilde{\lambda}^{2}_{1}+12}-\tilde{\lambda}_{1}]^{2}}{12}$ \\
(e1) &$-1$ &0&$1$&$1$  &$1$ & $1$ \\
(e2) &$1$ &0&$1$&$1$  &$1$ & $1$ \\
(e3) &$\sqrt{\frac{|\alpha|}{3}}$ &0&$1$&$\frac{|\alpha|}{3}$ & $1$ &$\frac{|\alpha|}{3}$ \\
(e4) &$-\sqrt{\frac{|\alpha|}{3}}$ &0&$1$&$\frac{|\alpha|}{3}$ & $1$ &$\frac{|\alpha|}{3}$ \\
\end{tabular}
\end{ruledtabular}
\end{table*}
The fixed points (d1)- (d4) are ``ultrarelativistic'' since the Lorentz factor $\gamma$ in 
Eq. \eqref{eq03} tends to infinity and thus $\tilde{\gamma}=0$. The other four fixed points (e1)-(e4) 
are ``nonrelativistic'' since the Lorentz factor $\gamma$ equals $1$ 
and thus $\tilde{\gamma}=1$. For these four ``nonrelativistic'' fixed points the DBI field 
mimics the behavior of a canonical scalar field.    

\subsubsection{{\bf Stability of the fixed points in Model II}}  

The stability analysis of the eight fixed points of Model II follows the same procedure as for 
Model I. For each of the eight fixed points listed in Table \ref{tab03} we found the
eigenvalues $\mu_{1}$ and $\mu_{2}$ of the matrix $\mathcal{M}$ in Eq. \eqref{eq26}. The results
for each point are given below along with whether the point is stable, unstable, or a saddle point. 
\begin{itemize}
  \item Point (d1):\\
          $\mu_{1}=\frac{3}{2}$, $\mu_{2}=\alpha$. \\
          This point is either a saddle point (if $\alpha <0$) or unstable (if $\alpha > 0$).
  \item Point (d2):\\
          $\mu_{1}=\frac{3}{2}$, $\mu_{2}=\alpha$. \\
          This point is either a saddle point (if $\alpha <0$) or unstable (if $\alpha > 0$).
  \item Point (d3):\\
$\mu_{1}=\frac{1}{4}\left(\tilde{\lambda}_{1}\left(\sqrt{\tilde{\lambda}^{2}_{1}+12}
-\tilde{\lambda}_{1}\right)-12\right)$,\\ 
$\mu_{2}=\frac{1}{2}\tilde{\lambda}_{1}\left(\sqrt{\tilde{\lambda}^{2}_{1}+12}-\tilde{\lambda}_{1}\right)+\alpha-3$. \\
This point is either stable [if  $\alpha<\frac{1}{2}(6+\tilde{\lambda}^{2}_{1})
-\frac{1}{2}\sqrt{\frac{\alpha^{2}-6\alpha+9}{\alpha}}$ ~] or unstable [if $\alpha >\frac{1}{2}(6+\tilde{\lambda}^{2}_{1})
-\frac{1}{2}\sqrt{\frac{\alpha^{2}-6\alpha+9}{\alpha}}$ ~] \\
 \item Point (d4):\\
           $\mu_{1}=\frac{1}{4}\left(\tilde{\lambda}_{1}\left(\sqrt{\tilde{\lambda}^{2}_{1}+12}
-\tilde{\lambda}_{1}\right)-12\right)$,\\
$\mu_{2}=\frac{1}{2}\tilde{\lambda}_{1}\left(\sqrt{\tilde{\lambda}^{2}_{1}+12}-\tilde{\lambda}_{1}\right)+\alpha-3$. \\
This point is either stable [if  $\alpha<\frac{1}{2}(6+\tilde{\lambda}^{2}_{1})
-\frac{1}{2}\sqrt{\frac{\alpha^{2}-6\alpha+9}{\alpha}}$ ~] or unstable [if $\alpha >\frac{1}{2}(6+\tilde{\lambda}^{2}_{1})
-\frac{1}{2}\sqrt{\frac{\alpha^{2}-6\alpha+9}{\alpha}}$ ~] \\
  \item Point (e1):\\
           $\mu_{1}=3$, $\mu_{2}=\alpha+3$. \\
           This point is either a saddle point (if $\alpha <-3$) or unstable (if $\alpha >-3$). \\
 \item Point (e2):\\
          $\mu_{1}=3$, $\mu_{2}=\alpha+3$. \\
          This point is either a saddle point (if $\alpha <-3$) or unstable (if $\alpha >-3$). \\
  \item Point (e3):\\
          $\mu_{1}=-\alpha-3$, $\mu_{2}=\frac{-3-\alpha}{2}$. \\
           This point is a stable (if $\alpha > 3$ ),  a saddle point (if $-3\leq \alpha \leq3$) or unstable (if $\alpha <-3$).\\
 \item Point (e4):\\
          $\mu_{1}=-\alpha-3$, $\mu_{2}=\frac{-3-\alpha}{2}$. \\
          This point is a stable (if $\alpha > 3$ ),  a saddle point (if $-3\leq \alpha \leq3$) or unstable (if $\alpha <-3$). \\
\end{itemize}

\begin{table*}
\caption{\label{tab04} The conditions for stability, acceleration, and existence for the eight 
critical points of Model II. We list the character of the fixed points as a function of the
parameters $\alpha$ and $\tilde{\lambda}_{1}$.} 
\begin{ruledtabular}
\begin{tabular}{llll}
Fixed point &Stability &Acceleration&Existence\\
\hline
(d1)&Saddle point for $\alpha<0$ &No&for $\alpha>0$ and $\alpha<0$ \\
{}    &Unstable node for $\alpha>0$ &{}&{}\\
(d2)&Saddle point for $\alpha<0$ &No&for $\alpha>0$ and $\alpha<0$\\
{}    &Unstable node for $\alpha>0$ &{}&{}\\
(d3) &Stable node for  $\alpha<\frac{1}{2}(6+\tilde{\lambda}^{2}_{1})
-\frac{1}{2}\sqrt{\frac{\alpha^{2}-6\alpha+9}{\alpha}}$ &$\tilde{\lambda}_{1}<2 $ &for $\alpha>0$ and $\alpha<0$\\
{}    &Unstable node for $\alpha>\frac{1}{2}(6+\tilde{\lambda}^{2}_{1})
-\frac{1}{2}\sqrt{\frac{\alpha^{2}-6\alpha+9}{\alpha}}$ &{}&{}\\
(d4) &Stable point for  $\alpha<\frac{1}{2}(6+\tilde{\lambda}^{2}_{1})
-\frac{1}{2}\sqrt{\frac{\alpha^{2}-6\alpha+9}{\alpha}}$ &$\tilde{\lambda}_{1}<2$ &for $\alpha>0$ and $\alpha<0$\\
{}    & Unstable node for
 $\alpha>\frac{1}{2}(6+\tilde{\lambda}^{2}_{1})-\frac{1}{2}\sqrt{\frac{\alpha^{2}-6\alpha+9}{\alpha}}$&{}&{}\\
(e1) &Saddle point for $\alpha<-3$ &No&for $\alpha>0$ and $\alpha<0$\\
{}    &Unstable node for $\alpha>-3$ &{}&{}\\
(e2) &Stable point for $\alpha<-3$  &No&for $\alpha>0$ and $\alpha<0$\\
{}    &Unstable node for  $\alpha>-3$ &{}&{}\\
(e3) &Stable node for $\alpha > 3$ & No&for $\alpha<0$\\
{}    &Saddle point for$-3\leq \alpha \leq3$ &{}&{}\\
{}    &Unstable node for  $\alpha<-3$ &{}&{}\\
(e4) &Stable node for $\alpha > 3$ &No&for $\alpha<0$\\
{}    &Saddle point for $-3\leq \alpha \leq3$ &{}&{}\\
{}    &Unstable node for  $\alpha<-3$ &{}&{}\\
\end{tabular}
\end{ruledtabular}
\end{table*}

\subsubsection{{\bf Late-time behavior of Model II}}

In this subsection we move on to the analysis of the late-time attractor structure of Model II
as given by the autonomous system in Eqs. \eqref{eq30} - \eqref{eq31}. The results for the eight 
fixed points of Model II are summarized in Tables \ref{tab03} and \ref{tab04}. The behavior 
of the dynamics of the DBI scalar field interacting with dark matter via $Q=\alpha H\rho_{m}$
depends on the values of the parameters $\alpha$ and $\tilde{\lambda}_{1}$. We found that there are
nontrivial ``scaling solutions'' where $x$, $y$, and $\tilde{\gamma}$ are finite constants depending on the 
model parameters  $\alpha$ and $\tilde{\lambda}_{1}$.  

\begin{figure}
\includegraphics[width=0.5\textwidth]{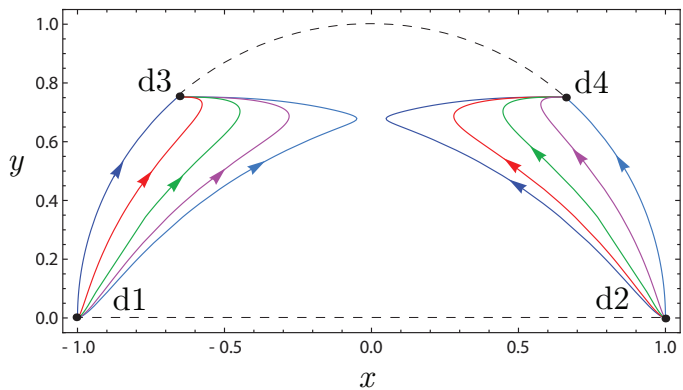}
\caption{\label{fig04} Model~II:  The phase plane for the DBI dark energy with interaction to dark matter 
corresponding to $\alpha=1$ for $\tilde{\lambda}_{1}=1$, $\tilde{\gamma}=0$. The
late-time attractors, (d3) and (d4), are stable fixed points which lead to accelerated expansion.}
\end{figure}

Tables \ref{tab03} and \ref{tab04}, show that in the interacting Model II, for ``ultrarelativistic" 
fixed points (d1) and (d2), $\mu_{1}$ is always positive, while $\mu_{2}$ can be 
either positive or negative depending on the value of $\alpha$. In particular, (d1) and (d2) 
are saddles point for $\alpha<0$ and are unstable for $\alpha>0$. For the fixed points (d3) and (d4) 
$\mu_{1}$ is always negative while $\mu_2$ can be positive or negative depending on the value of 
$\alpha$ and $\tilde{\lambda}_{1}$. In particular, (d3) and (d4) are stable, fixed points when
$\alpha<\frac{1}{2}(6+\tilde{\lambda}^{2}_{1})-\frac{1}{2}\sqrt{\frac{\alpha^{2}-6\alpha+9}{\alpha}}$ 
and are unstable points for 
$\alpha>\frac{1}{2}(6+\tilde{\lambda}^{2}_{1})-\frac{1}{2}\sqrt{\frac{\alpha^{2}-6\alpha+9}{\alpha}}$. 
These fixed points, (d3) and (d4), are the only
critical points from Model II which can give rise to an accelerated expansion for the Universe.
Accelerated expansion will occur for (d3) and (d4) if $\tilde{\lambda}_{1} < 2$.
Thus the fixed points (d3) and (d4) are good candidates for the late-time attractor solution under the
requirement that the parameters $\alpha$ and $\tilde{\lambda}_{1}$ are such that these points are stable
and that they meet the conditions for accelerated expansion. The $x-y$ phase-space flow of the points 
(d1) - (d4) are shown in Fig. \ref{fig04}. 

For the ``nonrelativistic" (i.e., $\tilde{\gamma}=1$) fixed points (e1) and (e2), 
$\mu_{1}$ is always positive, whereas $\mu_{2}$ can be either positive or negative
depending on the value of $\alpha$. If $\alpha<-3$ then $\mu_2<0$ and (e1) and (e2) are
saddle points. If $\alpha > -3$, then $\mu_2 > 0$ and (e1) and (e2) are unstable fixed points. 
For the points (e3) and (e4), $\mu_{1}$ and $\mu_{2}$ can be either positive or negative depending on the value of 
$\alpha$. These points are stable for $\alpha>3$, are saddle points for $-3\leq \alpha \leq 3$, and are unstable nodes 
for $\alpha<-3$. Thus none of the points (e1) - (e4) are stable, and from Table \ref{tab04} none of these
fixed points lead to accelerated expansion. Thus all of these points are not phenomenologically viable. 

Only points (d3) and (d4) satisfy the criteria of accelerated expansion (for values of $\tilde{\lambda}_{1} < 2$)
and stability (for certain values of $\alpha$ {\it and} $\tilde{\lambda}_{1}$). Here, in contrast to the case for
Model I, the coupling between dark energy and dark matter plays a significant role in the {\it stability}
of the fixed points. Although fixed points (d3) and (d4) would still exist without the coupling
between dark energy and dark matter characterized by the parameter $\alpha$, the stability of these points
does depend crucially on $\alpha$ and therefore on the coupling between dark energy and dark matter.  From Table
\ref{tab04} one can see that if one sets $\tilde{\lambda}_{1}=2$ (i.e., the maximum value for which one still gets
accelerated expansion), then $\alpha$ is restricted as
\begin{equation}
\label{alpha-cond}
2 \alpha + \sqrt{\frac{\alpha^2 -6 \alpha +9}{\alpha}} < 10.
\end{equation}
Solving this gives the restriction that
\begin{equation}
\label{alpha-cond-1}
0.088 < \alpha < 4.62 ~.
\end{equation}
Only for this range of $\alpha$ (and also one needs $\tilde{\lambda}_{1} < 2$) does one get both
a stable critical point and accelerated expansion. 

\section{Summary and Conclusions}
In this paper we studied the dynamics of dark energy (in the form of a Dirac-Born-Infeld
scalar field) interacting with dark matter (in the form of a fluid) in standard, flat FRW cosmology. 
For the scalar potential and the brane tension of the DBI scalar field we took power-law functions.
We analyzed two different models for the interaction between dark energy and dark matter.
Starting with the general interaction equation \eqref{eq04} and \eqref{eq05} we
considered $Q=\sqrt{2/3}\beta\rho_{m}\dot{\phi}$ (i.e., Model I) and $Q=\alpha H\rho_{m}$ (i.e., Model II).
For each model of the dark energy-dark matter interaction we were interested in fixed points 
which had accelerated expansion (i.e., $w_{eff} <-1/3$) {\it and} for which the fixed points were stable. 

For Model I, from Tables \ref{tab01} and \ref{tab02} one can see that points (a1), (a2), and
(b2) satisfy these two conditions. However, none of these points had any dependence on the coupling parameter
$\beta$ between the dark energy and dark matter. Thus for Model I there was no effect of adding the
coupling between dark energy and dark matter or a model which only had a DBI scalar field. Thus
while not ruled out, Model I is not really of interest since the coupling to dark matter does not
lead to any different result from simply having a DBI scalar field. 

For Model II we find two stable fixed points which have late time acceleration - points (d3) and (d4) - 
as can be seen in Tables \ref{tab03} and \ref{tab04}. In addition from Table \ref{tab04} one can see
that whether or not one has accelerated expansion depends on $\tilde{\lambda}_1$, which from Eqs. \eqref{eq14} 
and \eqref{eq15} depends on the scalar field potential and the tension, but does not depend on
the coupling between the DBI field and the dark matter fluid. Thus the addition of a coupling between dark
energy, in the form of a DBI scalar field and the dark matter fluid does not appear to contribute
to the existence of accelerated expansion for points (d3) and (d4) in Model II. A similar conclusion 
was reached for Model I where the stable fixed points with accelerated expansion did not depend
on the parameter $\beta$ which was a measure of the coupling between dark energy and dark matter for this
model. However, for Model II one can see, by looking at the ``stability" column of Table \ref{tab04}
that whether or not a given fixed point is stable {\it does} depend on $\alpha$ and therefore on the
coupling between the DBI field and the dark matter fluid. In particular in order for the critical points
(d3) and (d4) to lead to accelerated expansion {\it and} to be stable one needs to restrict the 
coupling parameter $\alpha$ between the DBI dark energy and dark matter fluid via Eq. \eqref{alpha-cond-1}.

Although we have only examined two specific types of couplings between dark energy (in the guise of a DBI
scalar field) and dark matter, we will tentatively advance some general conclusions about such models:

(i) The existence of fixed points with accelerated expansion does not depend on the coupling between the
dark energy and the dark matter. One can make a general argument to support this conclusion. Dark matter
will fall off like $[a(t)] ^{-3}$, while if the dark energy is to lead to late-time acceleration it will act
like a cosmological constant which falls off like $[a(t)] ^0 \rightarrow$ constant. At late times the 
$[a(t)] ^0$ behavior will always dominate the $[a(t)] ^{-3}$ behavior.

(ii) While the coupling to dark matter does not seem to play a role in the existence of fixed points with
accelerated expansion it can play a role in their stability. For example, in Model II the stability of the
fixed points (d3) and (d4), which had accelerated expansion, depended on the dark energy-dark matter coupling
parameter $\alpha$. However for Model I none of the fixed points' stability depended on $\beta$, the
coupling parameter for this model. Thus it seems that whether coupling between dark energy and dark matter is
important to the stability of the fixed points depends crucially on the type of coupling one chooses.

\begin{acknowledgments}
The work of NY has been supported by the Thailand Center of Excellence in Physics, the work of 
SS was supported by  a Russian Foundation for Basic Research grant No. 11-02-01162 and a Fulbright Scholars Grant,
and the work of DS is supported by a DAAD grant. We would like to thank Dr. Burin Gumjudpai for discussions 
and guidance. We also thank  Prof. Eduardo Guendelman, Prof. Edouard B. Manoukian, 
Dr. Piyabut Burikham, Dr. Khampee Karwan and Dr.Suppiya Siranan for discussions and comments. 
Finally, we would like to acknowledge with thanks for Department of Physics, California State University, 
Fresno, USA, for its kind hospitality.
\end{acknowledgments}

\end{document}